\documentclass[aps,twocolumn,prl,10pt,showpacs,byrevtex]{revtex4-1}
\usepackage{lineno}
\usepackage{times}
\usepackage{amsmath}
\usepackage{graphicx}
\usepackage{color}
\usepackage[T1]{fontenc}
\usepackage[implicit=true]{hyperref}
\usepackage{multirow}
\usepackage{bm}
\usepackage{amsmath}
\usepackage{xcolor}
\hypersetup{hidelinks}
\hypersetup{colorlinks = true, 
	    linkcolor = black, 
	    urlcolor = blue,
      citecolor = blue} 

\newcommand{\jpsi} {J/\psi}
\newcommand{\jlsb} {J/\psi \to \Lambda \bar{\Sigma}^{0} + c.c.}
\newcommand{\jllg} {J/\psi \to \Lambda \bar{\Lambda} \gamma}
\newcommand{\jll} {J/\psi \to \Lambda \Lambda + c.c.}
\newcommand{\jllb} {J/\psi \to \Lambda \bar{\Lambda} }
\newcommand{\llb} {\Lambda - \bar{\Lambda}}
\newcommand{\totjpsi} {(10087\pm44)\times 10^{6}$ $J/\psi}
\newcommand{\ecms} {$\sqrt{s}=3.097~\mathrm{GeV}$}
\newcommand{\rate} {$1.4\times 10^{-6}$}
\newcommand{\omass} {$2.1\times 10^{-18}~\mathrm{GeV}$}
\newcommand{\otime} {$3.1\times10^{-7}$~s}
\let\oldequation\equation
\let\oldendequation\endequation

\renewenvironment{equation}
  {\linenomathNonumbers\oldequation}
  {\oldendequation\endlinenomath}

\begin{document}

\title{Search for $ \Lambda - \bar{\Lambda} $~oscillation in $J/\psi\rightarrow\Lambda\bar{\Lambda}$~decay}

\author{
\begin{small}
\begin{center}
M.~Ablikim$^{1}$, M.~N.~Achasov$^{4,c}$, P.~Adlarson$^{76}$, O.~Afedulidis$^{3}$, X.~C.~Ai$^{81}$, R.~Aliberti$^{35}$, A.~Amoroso$^{75A,75C}$, Q.~An$^{72,58,a}$, Y.~Bai$^{57}$, O.~Bakina$^{36}$
% ...
\\
\vspace{0.2cm}
(BESIII Collaboration)\\
\vspace{0.2cm} {\it
$^{1}$ Institute of High Energy Physics, Beijing 100049, People's Republic of China\\
% ...
$^{3}$ Bochum  Ruhr-University, D-44780 Bochum, Germany\\
$^{4}$ Budker Institute of Nuclear Physics SB RAS (BINP), Novosibirsk 630090, Russia\\
% ...
$^{35}$ Johannes Gutenberg University of Mainz, Johann-Joachim-Becher-Weg 45, D-55099 Mainz, Germany\\
$^{36}$ Joint Institute for Nuclear Research, 141980 Dubna, Moscow region, Russia\\
% ...
$^{57}$ Southeast University, Nanjing 211100, People's Republic of China\\
$^{58}$ State Key Laboratory of Particle Detection and Electronics, Beijing 100049, Hefei 230026, People's Republic of China\\
% ...
$^{72}$ University of Science and Technology of China, Hefei 230026, People's Republic of China\\
% ...
$^{75}$ University of Turin and INFN, (A)University of Turin, I-10125, Turin, Italy; (B)University of Eastern Piedmont, I-15121, Alessandria, Italy; (C)INFN, I-10125, Turin, Italy\\
$^{76}$ Uppsala University, Box 516, SE-75120 Uppsala, Sweden\\
% ...
$^{81}$ Zhengzhou University, Zhengzhou 450001, People's Republic of China\\
\vspace{0.2cm}
$^{a}$ Deceased\\
% ...
$^{c}$ Also at the Novosibirsk State University, Novosibirsk, 630090, Russia\\
% ...
}\end{center}
\vspace{0.4cm}
\end{small}
}
\affiliation{}
\date{\today}

\begin{abstract}
Using $(10087\pm44)\times 10^{6}$ $J/\psi$~decays collected by the BESIII detector at the BEPCII collider, 
we search for baryon number violation via $\Lambda-\bar{\Lambda}$ oscillation in the decay $J/\psi \to \Lambda \bar{\Lambda}$.
No evidence for $\Lambda-\bar\Lambda$ oscillation is observed.
The upper limit on the time-integrated probability of $\Lambda-\bar{\Lambda}$~oscillation is estimated to be ~$1.4\times 10^{-6}$, 
corresponding to an oscillation parameter less than ~$2.1\times 10^{-18}~\mathrm{GeV}$~ at $90\%$~ confidence level.
\end{abstract}

\maketitle

\newpage

\section{INTRODUCTION}
The existence of baryon number violation is a fundamental question in particle physics~\cite{bnva}. 
This phenomenon plays a crucial role in explaining the observed matter-antimatter asymmetry in the universe. 
The asymmetry could be a result of three conditions pointed out originally by Sakharov~\cite{bnvc}: existence of Charge (C) or Charge-Parity (CP) violation; baryon number 
violating interactions; and the existence of out-of-thermal-equilibrium conditions in the early 
universe.
CP violation has been extensively studied in both
non-collider and collider experiments.
The violation of baryon number implies the instability of the proton 
and the atomic nucleus, which would occur on a time scale comparable to the lifetime of the universe~\cite{lou}.
In many theoretical models, baryon number is not a natural exact symmetry~\cite{c1,c2,c3,c4}.
For example, in some grand unified theories, protons can decay to light 
quarks in a variety of ways.
This mechanism simultaneously breaks the conservation of baryon number ($B$) 
and lepton number ($L$) while keeping their difference $B-L$ constant~\cite{pdg}.
Moreover, if neutrinos are Majorana particles with small masses~\cite{bnvg}, 
this would imply the presence of $\Delta(B-L)=2$ interactions,
thereby suggesting the existence of nucleon-antinucleon ($n - \bar{n}$) oscillation~\cite{bnvh}.
Many experiments have been carried out to search for $n - \bar{n}$ oscillation and the lower limit for the $n - \bar{n}$ oscillation time is currently 
$8.6 \times 10^{7}$~s with a $90\%$ confidence level~\cite{bnvd, bnve, bnvf}.
However, few results have been reported related to hyperons.
Recently, the BESIII Collaboration published experimental results on $\Lambda-\bar\Lambda$ oscillation 
using the decay $J/\psi\to p K^{-} \bar{\Lambda}+c.c$.
(Hereinafter, "$c.c.$" is used to indicate the inclusion of the charge-conjugate process for all relevant processes throughout the paper.)
The upper limit for the oscillation parameter was reported to be $\delta m_{\Lambda\bar{\Lambda}}$ < $3.8\times 10^{-18} \mathrm{GeV}$ at the $90\%$~ confidence level~\cite{deltE}. 

To investigate $\Lambda-\bar\Lambda$ oscillation,  we search for the presence of $\jll$ starting with the coherent production of $\Lambda\bar{\Lambda}$ pairs in the decay $\jllb$~\cite{llbtheory}.
The time evolution of $\Lambda - \bar{\Lambda}$ oscillation can be described by a Schrödinger-like 
equation 
\begin{equation}\label{gongshi1}
   i\frac{\partial}{\partial t}\dbinom{\Lambda(t)}{\bar{\Lambda}(t)}=M\dbinom{\Lambda(t)}{\bar{\Lambda}(t)},
\end{equation}
where $M$ is Hermitian matrix, defined as
\begin{equation}\label{juzhen1}
  M=\begin{pmatrix}
      m_{\Lambda}-\Delta E_{\Lambda} & \delta m_{\Lambda\bar{\Lambda}} \\
      \delta m_{\Lambda\bar{\Lambda}} & m_{\bar{\Lambda}}-\Delta E_{\bar{\Lambda}}
    \end{pmatrix},
\end{equation}
and $\delta m_{\Lambda\bar{\Lambda}}$ is the transition mass between the $\Lambda$ and $\bar{\Lambda}$ 
during oscillation, $m_{\Lambda}(m_{\bar{\Lambda}})$ is the nominal mass of the $\Lambda(\bar{\Lambda})$ baryon, 
and $\Delta E_{{\Lambda}/\bar{\Lambda}}$ is the energy split due to an external magnetic field. 
After considering the magnetic field (1.0~T) in the interaction region of the BESIII detector, 
the magnetic moment of the $\Lambda$ and the flight length of the $\Lambda$,
$\Delta E_{{\Lambda}/\bar{\Lambda}}$ is neglected in our case~\cite{deltE}.
The oscillation rate, 
$P(\bar{\Lambda},t)$, of generating a $\bar{\Lambda}$ with a beam of free $\Lambda$ after time $t$ can 
be written as
\begin{equation}\label{probability}
    P(\bar{\Lambda},t)=\sin^{2}(\delta m_{\Lambda\bar{\Lambda}}\cdot t)e^{-t/\tau_{\Lambda}},
\end{equation}
where $t$ is the time when the oscillation is observed, and $\tau_{\Lambda} = (2.632\pm 0.02)\times 10^{-10}$s ~is the mean lifetime of the $\Lambda$~\cite{pdg}.

The time-integrated oscillation probability of $\Lambda \to \bar{\Lambda}$  is given by
\begin{equation}\label{probability_2}
  P(\bar{\Lambda})=\frac{\int_{0}^{\infty} \sin^{2}(\delta m_{\Lambda\bar{\Lambda}}\cdot t)e^{-t/\tau_{\Lambda}} \,dt}{\int_{0}^{\infty} e^{-t/\tau_{\Lambda}} \,dt}.
\end{equation}

Then the oscillation parameter $\delta{m}_{\Lambda\bar{\Lambda}}$ can be deduced as
\begin{equation}\label{delatm2}
   \delta{m}_{\Lambda \bar{\Lambda}} = \sqrt{\frac{P(\Lambda)}{2 \tau_{\Lambda}^{2}}}.
\end{equation}

In this paper, using $\totjpsi$ decays~\cite{jpsinumber}, accumulated at the center-of-mass energy of \ecms ~with the BESIII detector,
we report the $\Lambda - \bar{\Lambda}$ oscillation parameter based on the $\jll$~process for the 
first time.

\section{ BESIII DETECTOR AND DATASET}
The BESIII detector~\cite{detector1} records symmetric $e^{+}e^{-}$ collisions provided by the BEPCII storage ring~\cite{detector2}
in the center-of-mass energy range from 1.85 to 4.95 $\mathrm{GeV}$, with a peak luminosity of 
$1.1\times 10^{33}~\mathrm{cm^{-2}s^{-1}}$ achieved at $\sqrt{s}=3.773~\mathrm{GeV}$.
BESIII has collected large data samples in this energy region~\cite{detector3}. The cylindrical core of the
BESIII detector covers 93$\%$ of the full solid angle and consists of a beam pipe, a helium-based multilayer
drift chamber (MDC), a plastic scintillator time-of-flight (TOF) system, and a CsI(Tl) electromagnetic calorimeter (EMC),
which are all enclosed in a superconducting solenoidal magnet providing a 1.0 $\mathrm{T}$ magnetic field.
The magnetic field was 0.9 $\mathrm{T}$ in 2012, which affects 10.8\% of the total $J/\psi$ data.
The solenoid is supported by an octagonal flux-return yoke with resistive plate counter muon identification modules interleaved with steel.
The charged-particle momentum resolution at $1~\mathrm{GeV}$ is $0.5\%$, and the $dE/dx$ resolution is $6\%$ for electrons from Bhabha scattering.
The EMC measures photon energies with a resolution of $2.5\%$($5\%$) at $1~\mathrm{GeV}$ in the barrel (end cap) region. 
The time resolution in the TOF barrel region is 68 ps, while that in the end cap region is 110 ps. The end cap TOF system was upgraded in 2015
using multigap resistive plate chamber technology, providing a time resolution of 60 ps~\cite{detector4}.

Simulated data samples produced with a {\sc geant4}-based~\cite{GEANT4:2002zbu} 
Monte Carlo (MC) package, which
includes the geometric description of the BESIII detector and the
detector response, are used to determine detection efficiencies
and to estimate backgrounds. 
The simulation models the beam
energy spread and initial state radiation in the $e^+e^-$
annihilations with the generator {\sc kkmc}~\cite{KKMC}.
The inclusive MC sample includes both the production of the $J/\psi$
resonance and the continuum processes incorporated in {\sc
kkmc}~\cite{KKMC}.  All particle decays are modelled with {\sc
evtgen}~\cite{BesEvtGen}~using branching fractions(BFs) 
either taken from the
Particle Data Group~(PDG)~\cite{pdg}, when available,
or otherwise estimated with {\sc lundcharm}~\cite{Lundcharm}.
Final state radiation~(FSR)
from charged final state particles is incorporated using the {\sc
photos} package~\cite{photo}.
For the signal process and oscillation process, we generate $1\times 10^{7}$ $\jllb$ MC samples and 
$1\times 10^{6}$ $\jll$ MC samples, respectively,
considering the polarization and decay parameters measured in Ref.~\cite{alphavalue}.
Additionally, exclusive MC samples for the main peaking background, $\jlsb$ and $\jllg$, 
are generated accordingly.

\section{EVENT SELECTION}
In this analysis, we fully reconstruct the $\Lambda$/$\bar \Lambda$ with $p\pi^-/\bar p\pi^+$ in 
the processes $\jllb$ and $\jll$.
We designate the events from the decay $\jllb$ as $right~sign$ (RS) events, and the ones from $\jll$ as $wrong~sign$ (WS) events.
Therefore, the final states in RS and WS decays are $p\bar{p}\pi^{+}\pi^{-}$ 
and $pp\pi^{-}\pi^{-}/\bar{p}\bar{p}\pi^{+}\pi^{+}$, respectively.

In the $\jllb$~process, charged tracks must be well reconstructed in the MDC, with $|\cos\theta| < 0.93$, where 
$\theta$ is defined with respect to the symmetry axis of the MDC, referred to as the $z$ axis.
Events with at least four selected charged tracks are retained for further analysis.
Based on MC simulation, the proton and pion from the $\Lambda$ decay are well separated kinematically.
The charged tracks with momentum less than $0.5~\mathrm{GeV}$ are taken as pion candidates,  while the other tracks are regarded as protons if their Particle Identification (PID) hypothesis of proton probability is higher than other particle hypotheses.
PID for charged tracks combines measurements of the energy deposited in the MDC~($dE/dx$) and the flight time in the TOF to form likelihoods $\mathcal{L}(h) (h=K, \pi, p)$ 
for each hadron $h$ hypothesis.

The $\Lambda/\bar \Lambda$ is reconstructed by looping over all $p\pi^-/\bar p \pi^+$ combinations, 
and a vertex fit is performed to each combination, ensuring that the
$p\pi^-/\bar p \pi^+$ particles should originate from common decay vertexes.
The position at which the $\Lambda$ decays into a proton and a pion is referred to as the decay vertex, while the point where the $\Lambda$ originated is referred to as the production vertex, determined by reconstructing the electron-positron interaction point using events from Bhabha scattering.
The mass windows for the $\Lambda$~and $\bar{\Lambda}$ selection are set to be within $\pm3\sigma$, 
where $\sigma = 1.8~\mathrm{MeV}$ is the mass resolution of the $\Lambda$~obtained from the fit to the invariant mass distribution of $p\pi^{-}$ in the
signal MC sample. 
All $\Lambda/\bar{\Lambda}$ candidates are retained for further analysis.

A four-constraint (4C) kinematic fit, imposing energy-momentum conservation, is performed under the 
hypothesis of $\jllb$ to suppress background and improve the kinematic resolution~\cite{Yan:2010zze}.
If more than one $\Lambda\bar{\Lambda}$ pair is found, the one with the minimum $\chi^{2}_{4\rm C}$ value is retained.
To improve the significance of the signal, the $\chi^{2}_{4\rm C}$ requirement is optimized using a 
figure-of-merit method: $\frac{S}{\sqrt{S+B}}$, 
where $S$ and $B$ are the numbers of signal and background events from
signal MC and inclusive MC samples, respectively.
The nominal criterion is set as $\chi^{2}_{4\rm C}$ $<50$.

In the decay $\jll$, the selection criteria are identical to those applied in 
the $\jllb$ process, with the exception that the $p / \pi^{-}$ ($\bar{p} / \pi^{+}$) candidate 
should not be shared among multiple reconstructed $\Lambda$ ($\bar{\Lambda}$) 
in an event.
The detection efficiencies of $\jllb$ and $\jll$ ~are determined to be 34.3\% and 34.4\% respectively.

\section{SIGNAL Extraction}
The MC sample for the WS channel was generated assuming complete $\llb$ oscillation, with the oscillation 
rate $P(\bar{\Lambda},t)$ set to 1. Consequently, only the decay rate of $\bar{\Lambda}$ was considered in the simulation. 
To match the actual detection efficiency, the oscillation effects of $\Lambda$, as described in Eq.\ref{probability}, should be 
convolved with the decay rate of $\bar{\Lambda}$ to construct a combined oscillation-decay rate distribution as a 
function of the  $\bar{\Lambda}$ decay vertex position. Using this distribution as the weight, the corrected detection 
efficiency for the WS channel is obtained by summing the re-weighted detection efficiency, which varies with 
the  $\bar{\Lambda}$ decay vertex position. 

Since the $P(\bar{\Lambda},t)$ is a function of the $\delta m_{\Lambda\bar{\Lambda}}$, one of the results of this measurement, an iterative process in the efficiency 
correction is necessary to account for the dependence on $\delta m_{\Lambda\bar{\Lambda}}$. However, simulation studies indicate that for $\delta m_{\Lambda\bar{\Lambda}}$ values 
below ${10}^{-15}$ GeV (which is significantly larger than the magnitude of the upper limit of ${10}^{-18}$ GeV), variations in $\delta m_{\Lambda\bar{\Lambda}}$ have 
a negligible effect on the distribution of $P(\bar{\Lambda},t)$. As a result, this simplifies the efficiency correction process, and does 
not affect the final results, including the upper limits. After applying the aforementioned efficiency correction, the corrected 
detection efficiency for the WS channel is found to be 14.9$\%$.

The background is estimated using the inclusive MC, exclusive MC, and data samples collected at 
$\sqrt{s}=3.773~\mathrm{GeV}$.
For the $\jllb$ process,
the ratio of the estimated background to the signal is about 0.2\% in the signal region,
including peaking background and non-peaking background.
The main peaking background sources are 
$\jpsi\rightarrow\bar{\Sigma}^{0}\Lambda+c.c$,
$\jpsi\rightarrow\Lambda\bar{\Lambda}\gamma$,
and 
$\jpsi\rightarrow\Lambda\bar{\Lambda}\to$
$\pi^+ \pi^- p \bar{p} \gamma  / 
e^- \bar{\nu}_{e} \pi^+ p \bar{p} / 
\mu^- \bar{\nu}_{\mu} \pi^+ p \bar{p} $
(
$\pi^+ \pi^- p \bar{p} \gamma / 
e^+ \nu_{e} \pi^- \bar{p} p / 
\mu^+ \nu_{\mu} \pi^- \bar{p} p $
).
These peaking backgrounds are simulated individually and normalized according
to the branching fractions taken from the PDG~\cite{pdg}. 
Only $605\pm27$ events survive, 
corresponding to 
$(2.00\pm0.09)\times 10^{-4}$
of the signal events in the data, which is treated as negligible.
With the data collected at $\sqrt{s}=3.773~\mathrm{GeV}$, the continuum 
contribution is studied. 
Assuming the cross section of the continuum process is proportional to $1/s$, 
the number of continuum background 
events, $N^{\rm cont.}_{3.097}$, is normalized to the $\jpsi$ center-of-mass energy using:
\begin{equation}\label{normal}
  N^{\rm cont.}_{3.097}=N^{\rm obs.}_{3.773}\times f_{\rm norm.},
\end{equation}
\begin{equation}\label{normal_factor}
  f_{\rm norm.}={\frac{\mathcal{L}_{3.097}}{\mathcal{L}_{3.773}}} \times{\frac{3.773^{2}}{3.097^{2}}}.
\end{equation}
Here $\mathcal{L}_{3.097}$ and $\mathcal{L}_{3.773}$ are the integrated luminosities 
of the data samples taken at $\sqrt{s}=3.097~\mathrm{GeV}$ and $3.773~\mathrm{GeV}$, respectively,
corresponding to 2962.7 pb$^{-1}$~\cite{lumjpsi} and 2931.8 pb$^{-1}$~\cite{lum3773},
and $f_{\rm norm.}$~is the normalization factor, which is calculated to be 1.5.
After the normalization, the number of continuum background events is calculated
to be $402\pm25$, corresponding to
$(1.3\pm0.8)\times 10^{-4}$
of the signal events and is also neglected.
For the process $\jll$, no events from either the inclusive MC or the data samples survive the event selection.

\begin{figure}[htbp]
  \centering
  \includegraphics[width=0.4\textwidth]{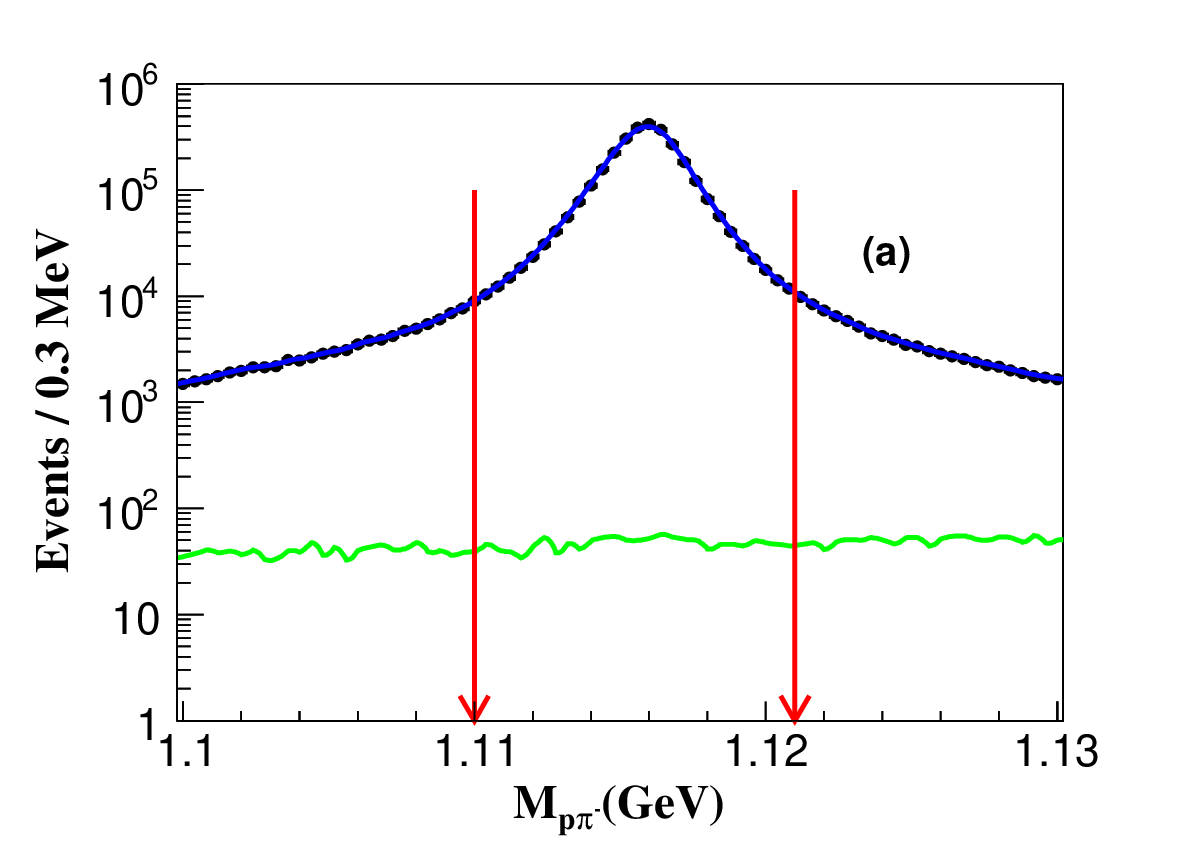}
  \includegraphics[width=0.4\textwidth]{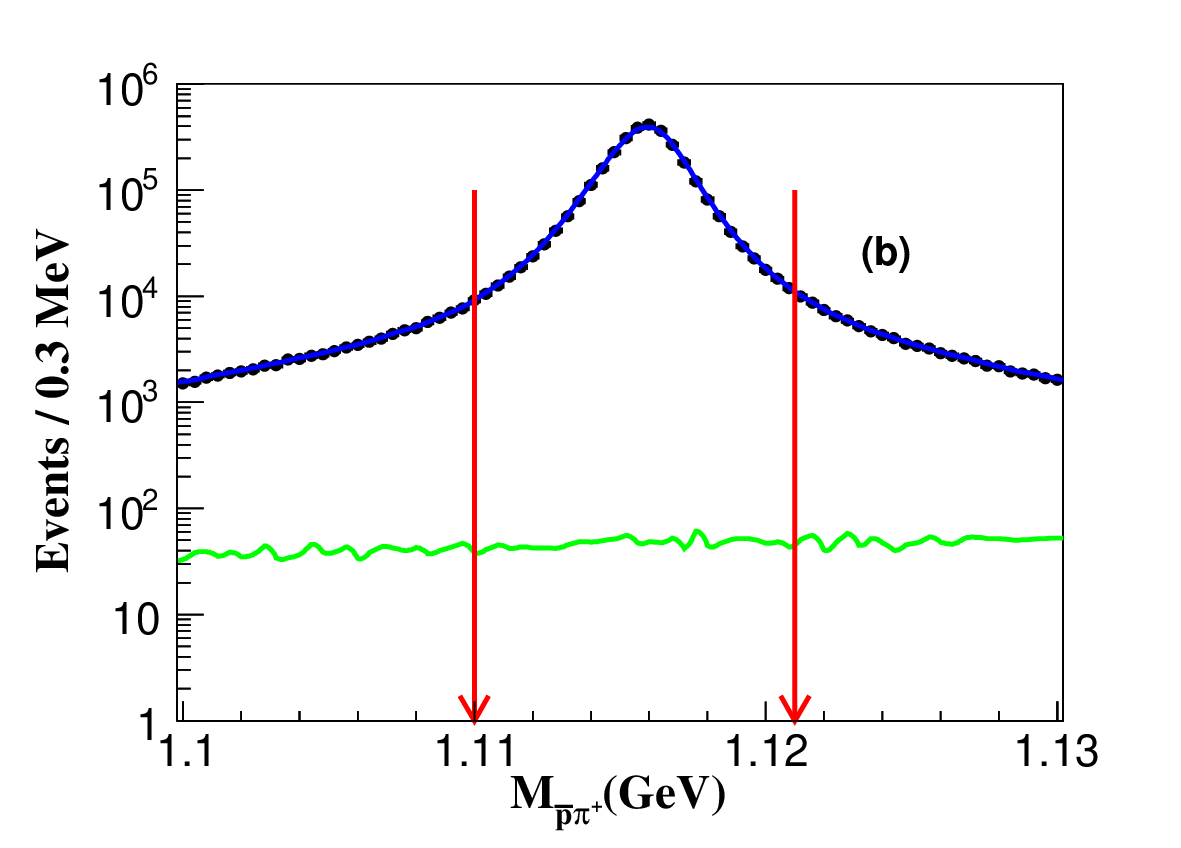}
  \caption{Fits to the mass distributions of (a) $M_{p\pi^{-}}$ and 
  (b) $M_{\bar{p}\pi^{+}}(\mathrm{GeV}$)
  in $J/\psi \to \Lambda \bar{\Lambda}$ in data, shown in log scale, where
  the filled circles with error bars are data, 
  the blue solid lines represent the results of the fits,
  the green solid lines show the background contributions,
  and the arrows in the figure show the edges of the signal region.}
\label{analysis:new_fitresult}
\end{figure}

A simultaneous fit of the $\Lambda$ and $\bar{\Lambda}$ mass spectrum is performed to determine the signal yield of $\jllb$.
The invariant mass $M_{p\pi^{-}}/M_{\bar{p}\pi^{+}}$ is obtained using the reconstructed momenta of 
the two charged tracks after passing all criteria.
The probability density functions of the signal distributions are derived from simulation, and are
convolved with a Gaussian function with floating width to account for the resolution difference between experimental data and MC simulation.
The background shape is described by the
parameterization of the non-peaking background simulation in the inclusive MC sample.
The fit range is $[1.1,1.13]~\mathrm{GeV}$ and the signal region is $[1.11,1.121]~\mathrm{GeV}$.
Figure~\ref{analysis:new_fitresult} shows the fit result.
The signal yield for the $\jllb$ decay is determined to be ~$3123264\pm1767$ events. 

\section{Upper limit Calculation}
Since no signal or background events are found in the search for $\jll$, 
a maximum likelihood estimator, based on the extended Profile
Likelihood method~\cite{trolke}, is constructed.
The numbers of signal and background events
are assumed to follow a Poisson distribution, and 
the detection efficiency is determined with an associated uncertainty, which is modeled as a Gaussian distribution. 
The likelihood is defined as
\begin{equation}\label{likelihood}
  \begin{split}
    \mathcal{L}=\mathcal{P}(\mathbf{N^{\rm UL}_{\rm WS}};N^{\rm obs}_{\rm WS}
    \cdot\boldsymbol{\epsilon_{\rm WS}}+\mathbf{N_{\rm bkg}})
    \otimes \mathcal{G}(\boldsymbol{\epsilon_{\rm WS}};\epsilon_{\rm WS},\sigma_{\epsilon_{\rm WS}})\\
    \otimes \mathcal{P}(\boldsymbol{N_{\rm bkg}};N_{\rm bkg}),
  \end{split}
\end{equation}
where $N^{\rm obs}_{\rm WS}$ and $N_{\rm bkg}$ are the observed number of signal events and background events in $\jll$ decay, respectively;
$\epsilon_{\rm WS}$ is the detection efficiency, and its
the standard deviation $\sigma_{\varepsilon_{\rm WS}}$ is derived from the systematic uncertainties that have already been canceled out of common sources,
which will be introduced later in the text.

The upper limit on the number of events, $\mathbf{N^{\rm UL}_{\rm WS}}$, for $\jll$,
is determined to be 13.0 at the 90 $\%$ confidence level. 
According to Eq.\ref{probability_2}, the probability of $\Lambda - \bar{\Lambda}$~oscillation in $\jllb$, denoted as $P(\Lambda)$,
can be expressed as 

\begin{equation}\label{sysratio}
  P(\Lambda) 
  = \frac{\mathcal{B}(J/\psi \to \Lambda \Lambda + c.c.)}{\mathcal{B}(J/\psi \to \Lambda \bar{\Lambda})}
  < \frac{\mathbf{N^{\rm UL}_{\rm WS}}}{N^{\mathrm{obs}}_{\mathrm{RS}}/\epsilon_{\mathrm{RS}}} = 1.4\times 10^{-6},
\end{equation}
where $\mathcal{B}(\jll)$ is shorthand for 
$\mathcal{B}(\jllb \overset{oscillating}{\rightarrow} \Lambda \Lambda +c.c.)$, 
representing the BF for the WS channel, and $\mathcal{B}(\jllb)$ is for the RS channel. 

As a result, the upper limit of the oscillation parameter $\delta{m}_{\Lambda \bar{\Lambda}}$ is calculated
to be less than \omass.

\section{systematic uncertainty}
In this measurement, the systematic uncertainties from the total number 
of $J/\psi$ events, tracking efficiencies, and PID efficiencies cancel.
The remaining systematic uncertainties are mainly from 
$\Lambda$ reconstruction, the signal shape, fit range, and kinematic fit.

The combined systematic uncertainty from $p$ and $\pi$~tracking efficiencies and the $\Lambda$ reconstruction 
efficiency has been estimated using a control sample of $J/\psi\rightarrow\bar{p}K^{+}\Lambda$.
This uncertainty is 0.6\% per reconstructed $\Lambda$~\cite{syserror2}. 

The uncertainty of the signal shape is estimated by using an alternative shape
modeled with a Breit-Wigner function convolved with a double Gaussian function.
The difference in the signal yield between the two approaches, 0.1\%, 
is taken as the uncertainty of the signal shape.

The uncertainty arising from the mass window regions is determined through a study of the resolutions of both data and MC, with a value of $0.1\%$.
Furthermore, the stability of the mass windows is studied by varying the range from 
$[1.110,1.121]~\mathrm{GeV}$~to $[1.109,1.120]~\mathrm{GeV}$~and $[1.111,1.122]~\mathrm{GeV}$.
The maximum deviation observed is 0.6\%.

\begin{figure}[htbp]
  \centering
  \includegraphics[width=0.4\textwidth]{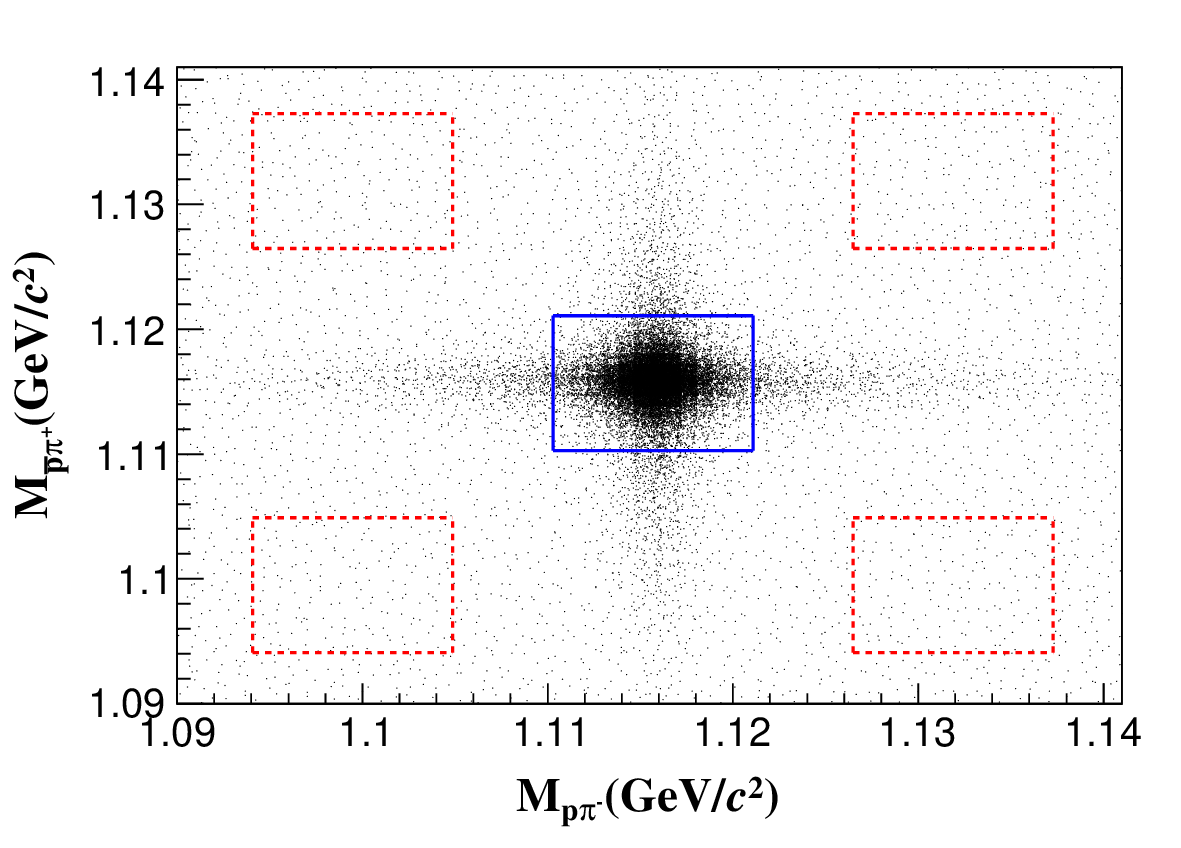}
  \caption{The 2D sideband scatter plot of the $\Lambda$ and $\bar{\Lambda}$ mass distributions from the $\jllb$ data sample, 
  where the box within the blue solid lines is the $\Lambda\bar{\Lambda}$ signal region, and the boxes bound by the red dashed lines are the sideband regions.}
  \label{sideband}
\end{figure}

The uncertainty due to unknown non-peaking backgrounds is estimated by replacing the number of 
background events obtained from the inclusive MC sample with the estimated number from data.
The estimated number of background events is obtained by using the 2-dimensional sideband method.
The sideband range is shown in Fig.~\ref{sideband}, and 
the average number of events within the four red boxes is calculated as the 
estimated number of background events. 
The final systematic uncertainty for the fit procedure is less than $0.1\%$, and can be safely neglected.

The uncertainty arising from the 4C kinematic fit
is estimated by using the track 
parameter correction method~\cite{BESIII:2012mpj}. The difference between 
the efficiencies with and without correction is taken as the systematic 
uncertainty. The correction factors for $p$ 
and $\bar{p}$ are determined based on a 
control sample of $\jpsi\rightarrow p K^{-}\bar{\Lambda}+c.c.$~and
$\psi(3686)\rightarrow p K^{-}\bar{\Lambda}+c.c.$, 
and the $\pi^{+}$ and $\pi^{-}$ are obtained based on the 
decay $\psi(3686)\rightarrow K^{+}K^{-}\pi^{+}\pi^{-}$ \cite{consample}.
The uncertainties caused by the kinematic fit for $\jllb$ and $\jll$~are all determined to be 0.7\%.

The uncertainty arising from misidentifying RS events as WS events 
is estimated using MC matching and found to be negligible.

The systematic uncertainties are listed in Table~\ref{summarysys}. 
Each source of systematic uncertainty is treated as independent
and they are summed in quadrature. 
\begin{table}[htbp]
  \centering
  \caption{Systematic uncertainties (in \%) in the measurement of each channel after cancellation of common sources.}
  \label{summarysys}
  \begin{tabular}{c c c}
  \hline\hline
       Source&$\jllb$&$\jll$\\ \hline
       $\Lambda$ reconstruction&  0.6 & 0.6     \\
       Kinematic fit &0.7  & 0.7  \\
       Mass windows  & 0.1 & 0.1 \\
       Signal shape & 0.1 & --  \\
       Mis-ID & ignore & ignore \\
       Bakground shape & ignore & -- \\ \hline
       Total & 0.9& 0.9 \\
       \hline\hline
  \end{tabular}
\end{table}

\section{summary}
In summary, based on $\totjpsi$ events at \ecms~ collected with the BESIII detector, 
a search for $\Lambda - \bar{\Lambda}$ oscillation is 
carried out in the decay $\jllb$, yielding no evidence. 
Consequently, upper limits at a 90\% confidence level are established for the time-integrated probability of
$\Lambda - \bar{\Lambda}$ oscillation and the oscillation parameter, set at $P(\Lambda)$ <~\rate 
and $\delta m_{\Lambda\bar{\Lambda}}$ <~\omass~respectively. 
This corresponds to an oscillation time limit of $\tau_{osc}$ > \otime~$(\tau_{osc}=1/\delta m_{\Lambda\bar{\Lambda}})$.
In comparison to the recent BESIII result~\cite{deltE}, $\delta m_{\Lambda\bar{\Lambda}}$ < $3.8\times 10^{-18}~\mathrm{GeV}$, 
we have imposed more stringent constraints on the upper limits of the corresponding parameters.

\section{Acknowledgement}
The BESIII collaboration thanks the staff of BEPCII and the IHEP computing center 
for their strong support. This work is supported in part by National Key R\&D Program of China
under Contracts Nos. 2023YFA1606000, 2020YFA0406300, 2020YFA0406400; Joint LargeScale Scientific Facility Funds of the NSFC and CAS
under Contract No. U1732102, U1832207; National Natural Science Foundation of China (NSFC) under Contracts Nos. 11635010, 11735014, 11935015, 11935016, 11935018, 12025502, 12035009,
12035013, 12061131003, 12192260, 12192261, 12192262, 12192263, 12192264, 12192265, 12221005, 12225509,
12235017, 12361141819, 12475082; the Chinese Academy of Sciences (CAS) Large-Scale Scientific Facility Program;
the CAS Center for Excellence in Particle Physics (CCEPP); 100 Talents Program of CAS; The Institute of Nuclear and Particle Physics (INPAC) and
Shanghai Key Laboratory for Particle Physics and Cosmology; The Natural Science Foundation of Shandong Province under Contract No. ZR2023MA004; 
German Research Foundation DFG under Contracts Nos. 455635585, FOR5327, GRK 2149; 
Istituto Nazionale di Fisica Nucleare, Italy; Ministry of Development of Turkey under
Contract No. DPT2006K-120470; National Research Foundation of Korea under Contract No. NRF-2022R1A2C1092335;
National Science and Technology fund of Mongolia; National Science Research and Innovation Fund (NSRF) via the
Program Management Unit for Human Resources \& Institutional Development, Research and Innovation of Thailand under
Contract No. B16F640076; Polish National Science Centre under Contract No. 2019/35/O/ST2/02907; The Swedish
Research Council; U. S. Department of Energy under Contract No. DE-FG02-05ER41374.


\end{document}